\documentclass[aps,prd,letterpaper,twocolumn,superscriptaddress,preprintnumbers,nofootinbib]{revtex4}

\usepackage{slashed}
\usepackage{amsmath,amssymb}
\usepackage{graphicx}
\usepackage{units}
\usepackage{bbold}
\usepackage{xcolor}
\usepackage{dsfont}
\usepackage[hyperfootnotes=false,colorlinks,citecolor=blue]{hyperref}

\usepackage{soul}

\usepackage{pifont}

\usepackage{comment}
\usepackage[normalem]{ulem}     

\usepackage{float}

\usepackage[caption=false,labelfont=normalsize,labelformat=empty,textfont=normalsize,justification=centering]{subfig}

\newcommand{\beq}{\begin{equation}}
\newcommand{\eeq}{\end{equation}}
\newcommand{\bea}{\begin{eqnarray}}
\newcommand{\ena}{\end{eqnarray}}

\newcommand{\ii}{{\rm i}}

\def \epsilon {\varepsilon}

\newcommand{\eq}{\mathrm{eq}}
\newcommand{\ie}{\emph{i.e.}}
\newcommand{\eg}{\emph{e.g.}}
\newcommand{\cf}{\emph{cf.}}

\newcommand{\D}{\mathrm{d}}
\newcommand{\E}{\mathrm{e}}

\newcommand{\TR}{T_\text{reh}}

\begin{document}

\title{Dirac leptogenesis from asymmetry wash-in via scatterings}

\preprint{TTK-24-14}

\author{Tom\'a\v{s} Bla\v{z}ek}
\email[E-mail: ]{tomas.blazek@fmph.uniba.sk}
\affiliation{Department of Theoretical Physics, Comenius University,\\ Mlynsk\'a dolina, 84248 Bratislava, Slovak Republic}

\author{Julian Heeck}
\email[E-mail: ]{heeck@virginia.edu}
\thanks{ORCID: \href{https://orcid.org/0000-0003-2653-5962}{0000-0003-2653-5962}.}
\affiliation{Department of Physics, University of Virginia,
Charlottesville, Virginia 22904-4714, USA}

\author{Jan Heisig}
\email[E-mail: ]{heisig@physik.rwth-aachen.de}
\affiliation{Institute for Theoretical Particle Physics and Cosmology, RWTH Aachen University,\\ D-52056 Aachen, Germany}
\affiliation{Department of Physics, University of Virginia,
Charlottesville, Virginia 22904-4714, USA}

\author{Peter Mat\'ak}
\email[E-mail: ]{peter.matak@fmph.uniba.sk}
\affiliation{Department of Theoretical Physics, Comenius University,\\ Mlynsk\'a dolina, 84248 Bratislava, Slovak Republic}

\author{Viktor Zaujec}
\email[E-mail: ]{viktor.zaujec@fmph.uniba.sk}
\affiliation{Department of Theoretical Physics, Comenius University,\\ Mlynsk\'a dolina, 84248 Bratislava, Slovak Republic}

\hypersetup{
pdftitle={Dirac leptogenesis and asymmetry wash-in via scatterings},
pdfauthor={Julian Heeck, Jan Heisig, Tom\'a\v{s} Bla\v{z}ek, Peter Mat\'ak, Viktor Zaujec}}

\begin{abstract}
Leptogenesis typically requires the introduction of heavy particles whose out-of-equilibrium decays are essential for generating a matter--antimatter asymmetry, according to one of Sakharov's conditions. We demonstrate that in Dirac leptogenesis, scatterings between the light degrees of freedom -- Standard Model particles plus Dirac neutrinos -- are sufficient to generate the asymmetry. The generation requires at least two effective charges conserved by the fast Standard Model interactions. Due to its vanishing source term in the Boltzmann equations, the asymmetry of right-handed neutrinos solely arises through wash-in processes.  Sakharov's conditions are satisfied because the right-handed neutrino partners are out of equilibrium. Consequently, heavy degrees of freedom never needed to be produced in the early universe, allowing for a reheating temperature well below their mass scale. Considering a minimal leptoquark model, we discuss the viable parameter space along with the potential observational signature of an increased number of effective neutrinos in the early universe.
\end{abstract}

\maketitle

\section{Introduction}
\label{sec:intro}

The discovery of nonzero neutrino masses through neutrino oscillations provides clear evidence of physics beyond the Standard Model (SM) and necessitates the introduction of new particles. Thus far, experiments have not resolved the fundamental property of neutrinos as either Dirac or Majorana particles, leaving two qualitatively different scenarios for their mass generation.

Majorana neutrinos, particularly when implemented in a seesaw mechanism, often offer suitable conditions for baryogenesis via leptogenesis~\cite{Fukugita:1986hr, Davidson:2008bu}. Essentially, the $\Delta L = 2$ interactions responsible for generating Majorana neutrino masses can induce a lepton asymmetry in the early universe, which can then be converted into a baryon asymmetry via sphalerons~\cite{Kuzmin:1985mm}. According to Sakharov's conditions~\cite{Sakharov:1967dj},
this process necessitates lepton number and $CP$ violation, as well as out-of-equilibrium dynamics, typically achieved through the freeze-out or freeze-in of the heavy seesaw states. The ability to simultaneously account for neutrino masses and the baryon asymmetry is a significant attraction of Majorana neutrino models.

However, Dirac neutrinos also have the potential to generate a matter-antimatter asymmetry, as demonstrated in Ref.~\cite{Dick:1999je}. In this scenario, the smallness of the Dirac-neutrino mass term effectively decouples the right-handed neutrino partners $\nu_R$ from the rest of the SM plasma. Without ever violating the lepton number, it becomes possible to create an effective lepton asymmetry by hiding an opposite asymmetry in the decoupled $\nu_R$ sector~\cite{Dick:1999je}. Similar to the seesaw case, fulfilling Sakharov's out-of-equilibrium condition typically involves the decay of heavy particles~\cite{Dick:1999je, Heeck:2023lxl}.

Interestingly, since the $\nu_R$ themselves are out of equilibrium, Dirac leptogenesis offers the possibility to satisfy Sakharov's conditions without actually involving the decay of heavy particles, $X_i$. As we show in this paper, it opens up a new variant of the mechanism in which the temperature of our universe never reached the heavy particle's mass scale while generating the asymmetry through freeze-in due to $2\to 2$ scatterings of SM particles and $\nu_R$. 
However, $CPT$ invariance and unitarity constraints~\cite{Kolb:1979qa, Dolgov:1979mz, Hook:2011tk, Baldes:2014gca} introduce additional conditions on the viability of the simple Dirac-leptogenesis models outlined in Ref.~\cite{Heeck:2023lxl}. In particular, the existence of at least three different final states in the scattering of $\nu_R$ is required for a non-vanishing asymmetry, as present in the case of $X_i$ being a leptoquark, \ie~model \textit{c} in~\cite{Heeck:2023lxl}. While the source term for the $\nu_R$ asymmetry in the Boltzmann equation is still vanishing, an asymmetry can be produced via wash-in~\cite{Domcke:2020quw}, provided there exists at least one additional effectively conserved charge such as right-handed electron number. In this work, we restrict ourselves to temperatures above $T\sim \unit[3\times 10^{13}]{GeV}$, where QCD sphalerons, except for the top Yukawa, are inefficient in coupling quark chemical potentials, thereby ensuring the presence of such conditions. We demonstrate that the resulting wash-in contribution is large enough to explain the observed baryon asymmetry, $Y_{\Delta B} \simeq 0.9\times 10^{-10}$~\cite{Davidson:2008bu, Planck:2018vyg},  rendering Dirac leptogenesis via scatterings a viable scenario. To the best of our knowledge, the underlying mechanism -- a combination of wash-in and freeze-in of the nonequilibrium particle density -- has not been considered in the literature before.

The remainder of the paper is structured as follows. In Sec.~\ref{sec:model}, we introduce the leptoquark model used throughout the study. We compute the scattering asymmetry and discuss the Boltzmann equations in Secs.~\ref{sec:asymmetry} and \ref{sec:boltzmann}, respectively. The implications for the baryon asymmetry are discussed in Sec.~\ref{sec:implications}. We conclude in Sec.~\ref{sec:conclusions}.

\section{Simple model}
\label{sec:model}

The simple renormalizable Dirac-leptogenesis models considered in Ref.~\cite{Heeck:2023lxl} can successfully explain the observed baryon asymmetry when assuming that the universe was hot enough to thermally produce a significant number of on-shell heavy mediator particles. However, here we are interested in the alternative scenario of a universe whose temperature never reached the mediators' mass scale in its entire (post-inflationary) history.
In this case, on-shell mediator production from the thermal bath is highly suppressed by the high-energy tails of thermal distributions and is, hence, negligible.
Accordingly, we could equally well do away with them altogether and describe their effects with effective operators, similar to Ref.~\cite{Hamada:2015xva}.
However, we opt to keep the mediator particles as degrees of freedom in our theory to facilitate direct comparison with the mediator decay case from Ref.~\cite{Heeck:2023lxl}.

As a simple model that contains all ingredients for successful Dirac leptogenesis via scattering, we introduce several scalar particles $X_i \sim (\Vec{3},\Vec{1},-1/3)$, \ie~the same gauge quantum numbers as the right-handed down quark, or the $S_1$ leptoquark in the notation of Ref.~\cite{Dorsner:2016wpm}. The relevant Yukawa interactions take the form 
\begin{align}\label{eq:Lag}
\mathcal{L}=\bar{Q}^c F_i L\bar{X}_i + \bar{d}^c_R G_i \nu^{\vphantom{c}}_R\bar{X}_i + \bar{u}^c_R K_i e^{\vphantom{c}}_R\bar{X}_i +  \text{H.c.}\,,
\end{align}
where we assumed the $X_i$ to be mass eigenstates with masses $M_i$. Focusing on a minimal model allowing for a successful leptogenesis, we consider two copies of $X$, $i=1,2$, coupled to the first generation of fermions only. 
To simplify our discussion, we have also imposed a baryon-number symmetry that forbids the di-quark couplings $Q Q X$ and $u_R d_R X$, \cf~case \textit{c} among the renormalizable Dirac-leptogenesis models from Ref.~\cite{Heeck:2023lxl}. 
Upon assigning $(B-L)(X_i) = -2/3$, the above Lagrangian is $B-L$ conserving. 
We stay agnostic about whether this is a global symmetry (and thus at most weakly broken through quantum-gravity effects) or a gauge symmetry (either unbroken with tiny coupling or large St\"uckelberg mass~\cite{Heeck:2014zfa}, or even spontaneously broken by more than two units to protect the Dirac-neutrino nature~\cite{Heeck:2013rpa,Heeck:2013vha}).
With these choices, the $X_i$ do not mediate any processes violating baryon number, lepton number, or lepton flavor.
These interactions technically contribute to the Dirac-neutrino mass at one-loop order, naively $\mathcal{O}( F^* m_d G/16 \pi^2)$ with the down-quark mass $m_d$, to be absorbed by the Dirac-mass counterterm together with the divergent piece.

While $B-L$ is hence conserved over the entire history of our universe, $\nu_R$ number need not be, allowing sphalerons~\cite{Kuzmin:1985mm} -- which are blind to the $\nu_R$ -- to convert the matching asymmetry $Y_{\Delta \nu_R}=Y_{\Delta (B-L_\text{SM})}$ into a baryon asymmetry~\cite{Kuzmin:1985mm, Harvey:1990qw}
\begin{align}
    Y_{\Delta B} = \frac{28}{79}Y_{\Delta (B-L_\text{SM})}= \frac{28}{79}Y_{\Delta \nu_R} \,.
    \label{eq:sphalerons}
\end{align}
The interactions in ${\cal L}$ indeed break $\nu_R$ number if  $G$ and either $F$ or $K$ are nonzero, seemingly allowing for the production of a $CP$ asymmetry in $\nu_R$. As we will show below, the necessary condition for a $\nu_R$ asymmetry is actually more subtle and requires the simultaneous presence of all three couplings in ${\cal L}$.

\section{Scattering cross sections and asymmetries}
\label{sec:asymmetry}

Right-handed neutrino interactions enter the matter--antimatter asymmetry evolution through thermally averaged cross sections. These are computed using \cite{Gondolo:1990dk}
\begin{align}
\langle\sigma v\rangle_{12\rightarrow 34} &= \frac{g_1g_2}{n^\eq_1 n^\eq_2}
\bigg(\prod^4_{a=1}\int\frac{\D^3 p_a}{(2\pi)^3 2E_a}\bigg)\mathrm{e}^{-\frac{E_1+E_2}{T}}\\
&\quad \times (2\pi)^4\delta^{(4)}(p_1+p_2-p_3-p_4) \vert\bar{\mathcal{M}}\vert^2\vphantom{\bigg)}\nonumber\\
&=\frac{g_1 g_2 T}{32\pi^4 n^\eq_1 n^\eq_2}
\int\D s\, s^{3/2}\sigma(s)K_1\bigg(\frac{\sqrt{s}}{T}\bigg) \,,\nonumber
\end{align}
where we employ the Maxwell--Boltzmann densities and neglect the Pauli blocking factors as an approximation.\footnote{If this is not the case and one wishes to keep the quantum statistics, it is necessary to include analogous statistical factors to cutting rules and asymmetry calculation. Otherwise, the $CPT$ and unitarity constraints \cite{Kolb:1979qa, Dolgov:1979mz, Hook:2011tk, Baldes:2014gca} and, consequently, the Sakharov conditions \cite{Sakharov:1967dj}, will be violated.} We assume kinetic equilibrium for all particles, including slowly-produced right-handed neutrinos. As discussed in Ref.~\cite{DEramo:2023nzt}, this approximation may be reasonably accurate for freeze-in production via scatterings.

All fermions in our model are considered massless. Therefore, in equilibrium, their densities $n^\eq_a$ can be written as a product of the right-handed neutrino equilibrium density $n^\eq_{\nu_R}=3\zeta(3)T^3/(4\pi^2)$ and the number of degrees of freedom $g_a$. 

The interactions are mediated by heavy $X_i$ scalars. In the $s$-channel, the cross sections contain singularities for $s=M^2_i$. These are usually regularized by inserting a finite mediator width, leading to double counting the on-shell $X_i$ production, which must be subtracted in the Boltzmann equation \cite{Kolb:1979qa}. Alternatively, one may follow the procedures introduced in Refs.~\cite{Ala-Mattinen:2023rbm, Matak:2023zox}. Focusing on temperatures much lower than the mediator masses, we neglect the energy dependence of the propagators, disregarding the region in which the singularities occur. Within this approximation, the lowest-order cross sections of the $t$- and $u$-channel processes are equal to one-third of the corresponding $s$-channel total cross sections. We will use this relation to simplify our Boltzmann equations.

For reactions containing two massless particles in the initial and final states, the $CPT$ symmetry allows us to define asymmetries of total cross sections as
\begin{align}
\Delta\sigma_{12\rightarrow 34}=
\sigma_{12\rightarrow 34}-
\sigma_{34\rightarrow 12}
\end{align}
obeying the unitarity constraints \cite{Kolb:1979qa, Dolgov:1979mz, Hook:2011tk, Baldes:2014gca}
\begin{align}\label{eq:unit}
\sum_f\Delta\sigma_{12\rightarrow f}=0\,,
\end{align}
where we sum over all possible final states. Relations of this type are essential to guarantee vanishing asymmetry in thermal equilibrium as required by Sakharov's conditions \cite{Sakharov:1967dj}. 

\begin{figure}
\subfloat{\label{fig1a}}
\subfloat{\label{fig1b}}
\subfloat{\label{fig1c}}
\subfloat{\label{fig1d}}
\centering\includegraphics[scale=1]{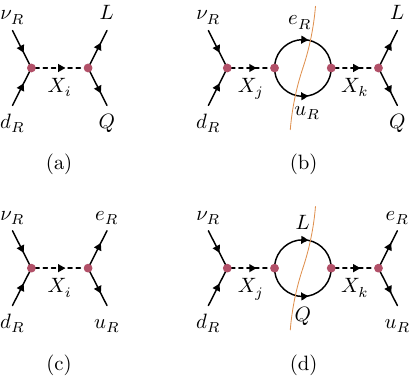}
\caption{\label{fig1} Tree and loop diagrams contributing to the $\nu_R d_R\rightarrow LQ$ and $\nu_R d_R\rightarrow e_R u_R$ asymmetries. The cuts represent the imaginary parts of the loop diagrams.}
\end{figure}

To see how unitarity and $CPT$ symmetry constrain the asymmetries in our model, let us look at the $\nu_R d_R\rightarrow LQ$ cross section. The asymmetry follows from the interference of the tree and loop diagrams in Figs.~\ref{fig1a} and \ref{fig1b}, respectively, and is proportional to
\begin{align}\label{eq:GFK}
\sum_{i,j,k}\frac{\Im[G^*_j G^{\vphantom{*}}_i F^*_i F^{\vphantom{*}}_k K^*_k K^{\vphantom{*}}_j]}{M^2_i M^2_j M^2_k}\, .
\end{align}
The couplings in the numerator of Eq.~\eqref{eq:GFK} can be represented by $2\times 2$ Hermitian matrices and parametrized as\footnote{A similar parametrization was used to study $CP$ violation in models containing multiple Higgs doublets \cite{Maniatis:2006jd, Maniatis:2006fs, Maniatis:2007de, Maniatis:2007vn, Nishi:2006tg, Ivanov:2006yq, Nishi:2007nh, Ivanov:2007de, Degee:2009vp, Ferreira:2023dke}.}
\begin{align}\label{eq:vec}
F^*_i F^{\vphantom{*}}_j = \frac{M_iM_j}{\TR^2}\sum_a f_a(\sigma_a)_{ij} \,,
\end{align}
where $f_a=(f_0,\Vec{f})$ are real dimensionless components of a Euclidean four-vector, $\sigma_a=(1,\Vec{\sigma})$, and $\Vec{\sigma}$ is the vector of Pauli matrices. 
In the single-flavor limit employed here, $f_0 = |\Vec{f}|$; furthermore, since $\TR\ll M_i$ by assumption, $|f_a|\ll 1$ in the perturbative-coupling regime. Analogously rewriting $G^*_i G^{\vphantom{*}}_j$ and $K^*_i K^{\vphantom{*}}_j$, thermal averaging yields the asymmetry
\begin{align}\label{eq:asym}
\Delta\langle\sigma_1 v\rangle= 
\frac{64}{\pi^2}\frac{T^4}{\TR^6}\frac{\Vec{g}.(\Vec{f}\times\Vec{k})}{\zeta(3)^2}
\equiv\frac{64}{\pi^2}\frac{T^4}{\TR^6}\frac{\epsilon}{\zeta(3)^2}\,,
\end{align}
where the subscript $1$ refers to the $\nu_R d_R\rightarrow LQ$ reaction; the cross sections of the other $s$-channel processes, $\nu_R d_R\rightarrow e_R u_R$ and $e_R u_R\rightarrow LQ$, will be further labeled by $2$ and $3$, respectively. 
Here, we introduced $\epsilon$ as a convenient dimensionless measure of $CP$ violation, further discussed below.

From the diagrams in Figs.~\ref{fig1c} and \ref{fig1d} we obtain the asymmetry in the $\nu_R d_R\rightarrow e_R u_R$ cross section proportional to $\Vec{g}. (\Vec{k}\times\Vec{f})=-\epsilon$ and clearly
\begin{align}\label{eq:unitcsv}
\Delta\langle\sigma_1 v\rangle+\Delta\langle\sigma_2 v\rangle=0\,.
\end{align}
For the $\nu_R d_R$ initial state, the only two-particle final states allowed by the Lagrangian density are $LQ$, $e_Ru_R$, and $\nu_Rd_R$. The total-cross-section asymmetry of the elastic scattering must vanish by the $CPT$ symmetry. Therefore, from the three possible final states mentioned in Sec.~\ref{sec:intro}, only two have nonzero asymmetries. These must be of opposite sign and cancel each other due to the unitarity constraints, as we observe in Eq.~\eqref{eq:unitcsv}.\footnote{In more complex models, the $CPT$ and unitarity constraints can be made explicit using the cyclic diagrams of Ref.~\cite{Roulet:1997xa} or holomorphic cutting rules \cite{Blazek:2021olf, Blazek:2021zoj, Blazek:2022azr}.}

Finally, the tree-level symmetric parts of the $s$-channel cross sections result in thermal averages
\begin{align}
\langle\sigma_1 v\rangle &= \frac{16}{3\pi}\frac{T^2}{\TR^4} \frac{f_0g_0+\Vec{f}.\Vec{g}}{\zeta(3)^2}\equiv\frac{16}{3\pi}\frac{T^2}{\TR^4}\frac{\alpha_1}{\zeta(3)^2}\,, \label{eq:csw1}\\
\langle\sigma_2 v\rangle &= \frac{8}{3\pi}\frac{T^2}{\TR^4} \frac{k_0g_0+\Vec{k}.\Vec{g}}{\zeta(3)^2}\equiv\frac{8}{3\pi}\frac{T^2}{\TR^4}\frac{\alpha_2}{\zeta(3)^2}\,,
\label{eq:csw2}\\
\langle\sigma_3 v\rangle &= \frac{16}{3\pi}\frac{T^2}{\TR^4} \frac{f_0k_0+\Vec{f}.\Vec{k}}{\zeta(3)^2}\equiv\frac{16}{3\pi}\frac{T^2}{\TR^4}\frac{\alpha_3}{\zeta(3)^2}\,,
\label{eq:csw3}
\end{align}
where we introduced $\alpha_{1,2,3}\geq 0$ as useful measures of cross-section strength. In the single-flavor limit, we have the inequality
\begin{align}\label{eq:epsilon}
\vert\epsilon\vert\leq\sqrt{2\alpha_1\alpha_2\alpha_3}
\end{align}
and our small-asymmetry calculation requires
\begin{align}\label{eq:epsilonalpha}
\vert\epsilon\vert\ll\frac{\pi}{12}\alpha_1,\frac{\pi}{24}\alpha_2,\frac{\pi}{12}\alpha_3\,.
\end{align}
The $\alpha_i$ are essentially free parameters, allowing, in particular, for large hierarchies.

\section{Boltzmann equations}
\label{sec:boltzmann}

To study particle densities and asymmetries, we follow the conventions of Ref.~\cite{Heeck:2023lxl} and introduce 
\begin{align}
\Sigma_a=\frac{n_a+n_{\bar{a}}}{s}\,, \quad\Delta_a=\frac{n_a-n_{\bar{a}}}{s}\,,
\end{align}
where $s = h_* T^3 \,2\pi^2/45$ denotes the entropy density. As an approximation, we assume the SM particles are in thermal equilibrium immediately after reheating and set the effective number of relativistic degrees of freedom entering the energy and entropy density to constant $g_{*}=h_{*}=106.75$. This assumption is apparently unrealistic but serves as a benchmark scenario to showcase the relevant dynamics. A more realistic reheating phase is expected to introduce $\mathcal{O}(1)$ corrections but not to change the qualitative picture.

\begin{figure*}[ht!]
\subfloat{\label{fig2a}}
\subfloat{\label{fig2b}}
\subfloat{\label{fig2c}}
\centering\includegraphics[scale=1]{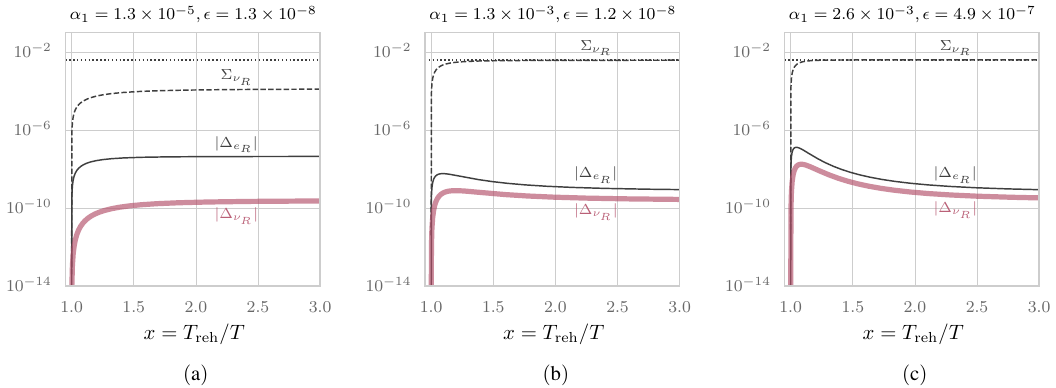}
\caption{\label{fig2} Evolution of $\Sigma_{\nu_R}(x)$ (black dashed line), $\Delta_{e_R}(x)$ (black solid line), and $\Delta_{\nu_R}(x)$ (thick red line) for multiple parameter values and $\TR=\unit[10^{14}]{GeV}$. The horizontal dotted line corresponds to $\Sigma^\eq_{\nu_R}\simeq 4\times 10^{-3}$. 
In all three cases, the ratio of the wash-out cross sections was fixed by $\alpha_2=2\alpha_1$ and $\alpha_3=\alpha_1$, and $\epsilon$ was chosen to obtain the observed baryon asymmetry.
}
\end{figure*}

As the initial abundance of right-handed neutrinos is negligible by assumption, they are produced via $LQ\rightarrow\nu_Rd_R$, $e_Ru_R\rightarrow\nu_Rd_R$, and crossed reactions. The symmetric part of their density obeys the Boltzmann equation
\begin{align}\label{eq:rhn}
\frac{\D\Sigma_{\nu_R}}{\D x}=-\frac{1}{x^4}\frac{\Gamma}{\mathcal{H}}\bigg\vert_{\TR} 
\bigg(\Sigma^{\vphantom{\eq}}_{\nu_R}-\Sigma^\eq_{\nu_R}\bigg)\,,
\end{align}
where $x\equiv \TR/T$, $\mathcal{H}= \pi \sqrt{g_*/90} \,T^2/M_\text{Pl}$ is the Hubble parameter with reduced Planck mass $M_\text{Pl}\simeq\unit[2.4\times 10^{18}]{GeV}$, and
\begin{align}\label{eq:gamma}
\Gamma=\frac{5}{2}s\Sigma^\eq_{\nu_R}(\langle\sigma_1 v\rangle+\langle\sigma_2 v\rangle)
\end{align}
parametrizes the $\nu_R$ interaction rate.
We solve Eq.~\eqref{eq:rhn} analytically and obtain
\begin{align}\label{eq:SigmaNu}
\Sigma_{\nu_R}(x)=\frac{135\zeta(3)}{4\pi^4 h_*}
\bigg(1-\exp\bigg[-\frac{\Gamma}{\mathcal{H}}\bigg\vert_{\TR}\frac{x^3-1}{3x^3}\bigg]\bigg)\,,
\end{align}
which rises steeply and quickly flattens out for $x\gg 1$, illustrated in Fig.~\ref{fig2}, reaching the equilibrium value for large interaction rates, $\Gamma/\mathcal{H}\vert_{\TR} \gg 1$, corresponding to 
\begin{align}
  2\alpha_1 + \alpha_2 \gg   \frac{\sqrt{g_*} \pi^4 \TR  \zeta(3)}{10\sqrt{10} M_\text{Pl}} \simeq\frac{\TR}{\unit[10^{17}]{GeV}} \,.
  \label{eq:thermalization}
\end{align}
Notice that the $\alpha_i$ themselves scale with $\TR^4$ times Lagrangian parameters.

The evolution of an asymmetry generally results from two competing parts. The source term contains asymmetries of thermally averaged cross sections, whereas the wash-out terms are proportional to particle density asymmetries $\Delta_a$. The latter are subject to constraints implied by the symmetries and conservation laws of the model. In particular, the interactions in Eq.~\eqref{eq:Lag} separately conserve the lepton and baryon numbers, resulting in
\begin{align}\label{eq:BL}
\Delta_{\nu_R}=-\Delta_{e_R}-\Delta_{L}\,,\,\quad
\Delta_{d_R}=-\Delta_{u_R}-\Delta_{Q}\,.
\end{align}
Moreover, there are three other global $U(1)$ symmetries: considering $\nu_R\rightarrow\E^{\ii\alpha}\nu_R$ with $d_R\rightarrow\E^{-\ii\alpha}d_R$ leaves the Lagrangian density invariant, and the same independently applies to $e_R$, $u_R$ and $Q$, $L$ field transformations. The conservation of the respective charges then implies
\begin{align}\label{eq:3U1}
\Delta_{\nu_R}=\Delta_{d_R}\,,\quad\Delta_{e_R}=\Delta_{u_R} \,,\quad
\Delta_L=\Delta_Q\,.
\end{align}
Eqs.~\eqref{eq:BL} and \eqref{eq:3U1} impose four relationships among the six asymmetries.
Therefore, it is sufficient to solve the Boltzmann equations for $\Delta_L$ and $\Delta_{e_R}$ asymmetries, while the others are fixed through Eqs.~\eqref{eq:BL} and \eqref{eq:3U1}. 

We note that these equations are only valid when neglecting SM Yukawa and sphaleron interactions. Therefore, we restrict ourselves to reheating temperatures above approximately $\unit[3\times 10^{13}]{GeV}$, where only the top-quark Yukawa and SM gauge interactions are fast compared to the Hubble rate.  Below $T\sim \unit[3\times 10^{13}]{GeV}$,  QCD sphalerons become efficient in relating the quark chemical potentials \cite{Bento:2003jv, Garbrecht:2014kda}. Above this temperature, the interactions in Eq.~\eqref{eq:Lag} generate asymmetries in charges of the first-generation fermions which are transformed into a baryon asymmetry according to Eq.~\eqref{eq:sphalerons} as the temperature decreases. For lower reheating temperatures, 
one has to study the evolution of asymmetries in charges effectively conserved by the interactions with rates much faster than the universe expansion \cite{Fong:2015vna, Fong:2020fwk}. In particular, below roughly $T\sim \unit[10^{4}]{GeV}$, besides the $\nu_R$ number, the only effectively conserved charge is $(B-L)_{\text{SM}}$ \cite{Buchmuller:2000as, Buchmuller:2005eh}. As its source term vanishes (see below), no asymmetry can be produced. For higher temperatures, on the other hand, the $e_R$ number will be preserved due to the small Yukawa couplings \cite{Buchmuller:2000as, Buchmuller:2005eh}. As it is sourced by asymmetric right-handed neutrino scatterings, our mechanism will work up to corrections of fast SM processes.

It has been argued in Ref.~\cite{Racker:2018tzw} that to calculate the source term of an asymmetry, correspondingly charged particles must be produced in contributing reactions, while in agreement with Sakharov's conditions \cite{Sakharov:1967dj}, the initial states must be out of equilibrium. In our scenario, the only out-of-equilibrium particles present in the universe are right-handed neutrinos. Once they are in the initial state, for the model in Eq.~\eqref{eq:Lag}, they cannot be produced in the same $2\rightarrow 2$ reaction. The source term of the right-handed neutrino asymmetry thus vanishes, but, fortunately, it does not mean that no asymmetry in their density can be produced. To observe this, let us finally write the Boltzmann equations for the density asymmetries,
\begin{align}
\frac{\D\Delta_{L}}{\D x}=&\,
\frac{\Sigma^\eq_{\nu_R}}{3\mathcal{H}}\frac{\D s}{\D x}\bigg\{
\frac{3}{4}\Delta\langle\sigma_1 v\rangle\bigg(\Sigma^\eq_{\nu_R}-\Sigma^{\vphantom{\eq}}_{\nu_R}\bigg)\label{eq:DL}\\
&+\frac{5}{3}\langle\sigma_3 v\rangle
\bigg(\Delta_L-2\Delta_{e_R}\bigg)\nonumber\\
&+\frac{4}{3}\langle\sigma_1 v\rangle
\bigg[\bigg(\Delta_L-\frac{17} {8}\Delta_{\nu_R}\bigg)\nonumber\\
&+\frac{1}{4}\frac{\Sigma_{\nu_R}}{\Sigma^\eq_{\nu_R}}
\bigg(\Delta_L-\frac{3}{2}\Delta_{\nu_R}\bigg)\bigg]\bigg\}\,,\nonumber\\
\frac{\D\Delta_{e_R}}{\D x}=&\,
\frac{\Sigma^\eq_{\nu_R}}{3\mathcal{H}}\frac{\D s}{\D x}\bigg\{
\frac{3}{4}\Delta\langle\sigma_2 v\rangle\bigg(\Sigma^\eq_{\nu_R}-\Sigma^{\vphantom{\eq}}_{\nu_R}\bigg)\label{eq:DeRold}\\
&+\frac{5}{3}\langle\sigma_3 v\rangle
\bigg(2\Delta_{e_R}-\Delta_L\bigg)\nonumber\\
&+\frac{4}{3}\langle\sigma_2 v\rangle
\bigg[\bigg(2\Delta_{e_R}-\frac{17}{8}\Delta_{\nu_R}\bigg)\nonumber\\
&+\frac{1}{4}\frac{\Sigma_{\nu_R}}{\Sigma^\eq_{\nu_R}}
\bigg(2\Delta_{e_R}-\frac{3}{2}\Delta_{\nu_R} \bigg)\bigg]\bigg\}\,.\nonumber
\end{align}
The first lines on the right-hand sides of Eqs.~\eqref{eq:DL} and \eqref{eq:DeRold} correspond to the source terms containing the cross section asymmetries. According to Eq.~\eqref{eq:unitcsv}, they only differ in sign. The lines below contain the wash-out terms which are linear in the density asymmetries $\Delta_L$, $\Delta_{e_R}$, and $\Delta_{\nu_R}=-\Delta_L-\Delta_{e_R}$. Again, the contributions of $e_R u_R\rightarrow LQ$ and crossed reactions enter Eqs.~\eqref{eq:DL} and \eqref{eq:DeRold} with opposite signs. If there were no other contributions, we would obtain $\Delta_L=-\Delta_{e_R}$ and vanishing $\Delta_{\nu_R}$.

However, the remaining wash-out terms depend on different combinations of $\Delta_L$ and $\Delta_{e_R}$. According to  Eq.~\eqref{eq:BL}, this allows for a nonzero and, in fact, sufficiently large $\Delta_{\nu_R}$ which is washed-in \cite{Domcke:2020quw}, even though its source term vanishes identically.\footnote{A similar situation occurs in purely flavored leptogenesis, where the flavor-dependent wash-out terms lead to a non-zero asymmetry, even though the total lepton-number source-term vanishes~\cite{AristizabalSierra:2009bh}.}

Following these arguments, it is instructive to rewrite Eqs.~\eqref{eq:DL} and \eqref{eq:DeRold} in terms of $\Delta_{\nu_R}$ and $\Delta_{e_R}$,
\begin{align}
\frac{\D\Delta_{\nu_R}}{\D x}=&\,
\frac{\Sigma^\eq_{\nu_R}}{3\mathcal{H}}\frac{\D s}{\D x}\bigg\{ \frac{5}{6} \langle\sigma_1 v\rangle \bigg(5+\frac{\Sigma_{\nu_R}}{\Sigma^\eq_{\nu_R}} \bigg)\Delta_{\nu_R}\label{eq:DnR}\\
&+\frac{1}{6}\langle\sigma_2 v\rangle  \bigg( 17 + 3\frac{\Sigma_{\nu_R}}{\Sigma^\eq_{\nu_R}}\bigg)\Delta_{\nu_R}\nonumber\\
&+\frac{4}{3}\bigg(\langle\sigma_1 v\rangle-2\langle\sigma_2 v\rangle\bigg)\bigg(1+\frac{1}{4}\frac{\Sigma_{\nu_R}}{\Sigma^\eq_{\nu_R}}\bigg)
\Delta_{e_R}\bigg\}\,,\nonumber\\
\frac{\D\Delta_{e_R}}{\D x}=&\,
\frac{\Sigma^\eq_{\nu_R}}{3\mathcal{H}}\frac{\D s}{\D x}\bigg\{
\frac{3}{4}\Delta\langle\sigma_2 v\rangle\bigg(\Sigma^\eq_{\nu_R}-\Sigma^{\vphantom{\eq}}_{\nu_R}\bigg)\label{eq:DeR}\\
&+\frac{5}{3}\langle\sigma_3 v\rangle
\bigg(3\Delta_{e_R}+\Delta_{\nu_R}\bigg)\nonumber\\
&+\frac{4}{3}\langle\sigma_2 v\rangle
\bigg[\bigg(2\Delta_{e_R}-\frac{17}{8}\Delta_{\nu_R}\bigg)\nonumber\\
&+\frac{1}{4}\frac{\Sigma_{\nu_R}}{\Sigma^\eq_{\nu_R}}
\bigg(2\Delta_{e_R}-\frac{3}{2}\Delta_{\nu_R} \bigg)\bigg]\bigg\}\,.\nonumber
\end{align}
Although the $\Delta_{\nu_R}$ source term in Eq.~\eqref{eq:DnR} vanishes, a nonzero asymmetry can originate from $\Delta_{e_R}$. The only exception occurs for $\alpha_1=\alpha_2$, or $\langle\sigma_1 v\rangle=2\langle\sigma_2 v\rangle$, as in this case, the coefficient in front of $\Delta_{e_R}$ in Eq.~\eqref{eq:DnR} vanishes, erasing any $\nu_R$ asymmetry.

The Boltzmann equations only depend on the ratios $\alpha_i M_\text{Pl}/(\sqrt{g_*}\TR)$ and $\epsilon  M_\text{Pl}/(h_*\sqrt{g_*}\TR)$, so the dependence on $\TR$ (and  $M_\text{Pl}$, $g_*$, and $h_*$) can be completely scaled away.
Furthermore, the Boltzmann equations have been derived under the approximation of \textit{small} asymmetries, which immediately implies that all asymmetries are directly proportional to $\epsilon  M_\text{Pl}/(h_*\sqrt{g_*}\TR)$, in particular $\Delta_{\nu_R}\propto \epsilon  M_\text{Pl}/(h_*\sqrt{g_*}\TR)$. This effectively leaves the three $\alpha_i M_\text{Pl}/(\sqrt{g_*}\TR)$ as free parameters in the Boltzmann equations.

Numerical solutions for $\TR=\unit[10^{14}]{GeV}$ and various coupling values are shown in Fig.~\ref{fig2}. In each of the three cases, fast $\nu_R$ production can be observed at the very beginning. In Fig.~\ref{fig2a}, the solution corresponds to weak right-handed neutrino interactions with the SM bath. Their density never reaches the equilibrium value and freezes in at $x\simeq 1.5$. The same happens to $\vert\Delta_L\vert$ and $\vert\Delta_{e_R}\vert$, which are roughly the same, while $\Delta_{\nu_R}$ is suppressed by two orders of magnitude as it only comes from their tiny difference.

In Figs.~\ref{fig2b} and \ref{fig2c}, at stronger couplings, both source and wash-out cross sections are enhanced, and initially produced asymmetries are larger. However, when right-handed neutrinos reach equilibrium, the source term vanishes, and the asymmetries are partially washed out.

\section{Baryon asymmetry} 
\label{sec:implications}

\begin{figure*}
\subfloat{\label{fig3a}}
\subfloat{\label{fig3b}}
\centering\includegraphics[scale=1]{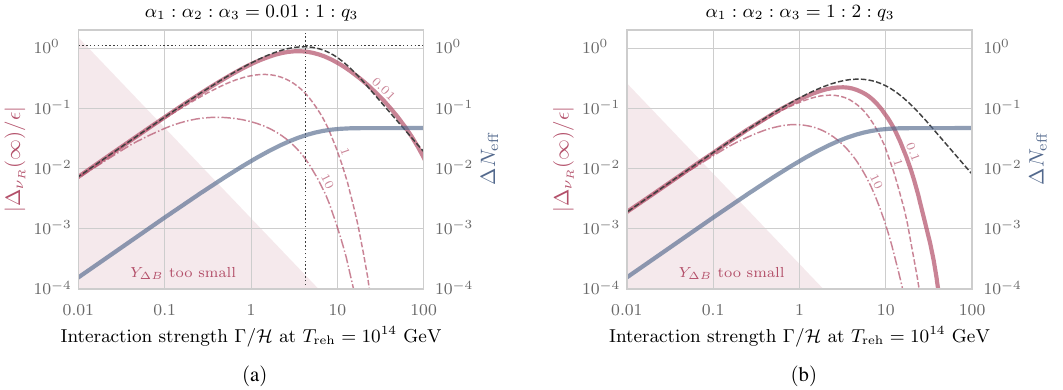}
\caption{\label{fig3} Numerically computed values of $\vert\Delta_{\nu_R}(\infty)/\epsilon\vert$ (red contours labeled by $q_3$) plotted with $\Delta N_{\text{eff}}$ (gray line) for $\TR=\unit[10^{14}]{GeV}$. The ratio of the wash-out cross sections was fixed as indicated in the upper part of each panel. 
The black dashed line shows the freeze-in approximation from Eq.~\eqref{eq:full_approx}.
In Fig.~\ref{fig3a}, the black dotted lines correspond to the maximal asymmetry from Eq.~\eqref{eq:maxasymm} and the respective $\alpha_2$ value from Eq.~\eqref{eq:alpha2}. The red-shaded region is excluded by Eq.~\eqref{eq:epsilon} for $q_3=0.01$; for higher $q_3$ values, this constraint will be slightly relaxed. 
}
\end{figure*}

While the differential equations for $\Delta_L$ and $\Delta_{e_R}$ are not analytically solvable in general, we can obtain an analytic approximation in the limit of negligible $\Sigma_{\nu_R}$, \ie~in the freeze-in regime:
\begin{align}
\Delta_{\nu_R}(\infty)&\simeq \frac{1296\,\zeta(3) (\alpha_1-\alpha_2)(\Gamma/\mathcal{H})\vert_{\TR}\, \epsilon }{\pi^5 h_*  (2\alpha_1+\alpha_2) (50\alpha_1+\alpha_2-60\alpha_3)} \nonumber\\
&\quad \times \left[ \xi \left(\frac{50\alpha_1+17\alpha_2}{90\alpha_1+45\alpha_2}\, \frac{\Gamma}{\mathcal{H}}\bigg\vert_{\TR}\right)\right. \label{eq:full_approx}\\
&\qquad \left.- \xi \left(\frac{16\alpha_2+60\alpha_3}{90\alpha_1+45\alpha_2}\, \frac{\Gamma}{\mathcal{H}}\bigg\vert_{\TR}\right) \right] , \nonumber 
\end{align}
where $\xi(y)$ is proportional to the incomplete Gamma function $\Gamma (5/3,0,y)$:
\begin{align}
    \xi (y ) \equiv \frac{\Gamma (5/3,0,y)}{y^{5/3}} = \frac{5}{3} - \frac{3y}{8} - \frac{3y^2}{22} + \mathcal{O}(y^3)\,.
\end{align}
Notice that $\Delta_{\nu_R}(\infty)$ is well behaved in the limit $50\alpha_1+\alpha_2-60\alpha_3\to 0$. At leading order in small $\alpha_i$, \ie~for small couplings in Eqs.~\eqref{eq:csw1}--\eqref{eq:csw3}, the expression simplifies considerably:
\begin{align}
\Delta_{\nu_R}(x\to\infty)&\simeq 0.028\,\frac{M_\text{Pl}(\alpha_2-\alpha_1)}{\sqrt{g_*}\TR}
 \frac{M_\text{Pl}}{\sqrt{g_*}h_*\TR}\epsilon  \label{eq:Delta_nuR_approx}\\
&\sim \frac{0.028}{g_*h_* }\frac{M^2_\text{Pl}}{\TR^2} \,\mathcal{O}\left(\frac{\TR^{10}}{M^{10}}F^{10}\right) ,
\end{align}
where in the second line we have assumed both $X$ masses to be of similar order and all Yukawa couplings of order $F$ to get a parametric dependence. 
Notice that in this freeze-in regime, $\alpha_3$ plays no role since the underlying reaction $e_R u_R \to L Q$ does not involve $\nu_R$; numerically, we find that increasing $\alpha_3$ for fixed $\epsilon$  always suppresses the final asymmetry, see e.g.~Fig.~\ref{fig3}.

Equation~\eqref{eq:Delta_nuR_approx} shows an increase in $\nu_R$ asymmetry for increasing $\alpha_{1,2}$, typical for the freeze-in regime. Once the interactions become strong enough to thermalize the $\nu_R$, Eq.~\eqref{eq:thermalization}, the asymmetry will of course start to exponentially decrease.
Still in the limit of weak $\nu_R$ interactions, we can approximate the \emph{maximal} $\nu_R$ asymmetry using Eq.~\eqref{eq:full_approx} as
\begin{align}\label{eq:maxasymm}
    \Delta_{\nu_R}^\text{max}(x\to\infty)&\simeq 0.05\,\frac{M_\text{Pl}}{\sqrt{g_*}h_*\TR}\epsilon \,,
\end{align}
which can be achieved for
\begin{align}\label{eq:alpha2}
    \alpha_{1,3}\ll \alpha_2 \simeq 5.1 \, \frac{\sqrt{g_*} \TR}{M_\text{Pl}}
\end{align}
or, almost, and with opposite sign, for values
\begin{align}
    \alpha_{2,3}\ll \alpha_1 \gg 5.1 \, \frac{\sqrt{g_*} \TR}{M_\text{Pl}}\,.
\end{align}
Even though the position and value of the $\Delta_{\nu_R}^\text{max}(x\to\infty)$ lie strictly speaking outside of the freeze-in regime used in our analytical approximation, they serve as useful guides. 

For fixed ratios of the three $\alpha_i$, we show the final $\nu_R$ asymmetry as a function of the $\nu_R$ interaction rate $\Gamma$ from Eq.~\eqref{eq:gamma} over Hubble at reheating in Fig.~\ref{fig3}. We can clearly see the features derived above: for small interaction rates, the linear freeze-in approximation from Eq.~\eqref{eq:Delta_nuR_approx} works very well, although deep into that regime the inequality from Eq.~\eqref{eq:epsilon} makes it impossible to obtain the measured baryon asymmetry. At higher reheating temperatures, the constraint of Eq.~\eqref{eq:epsilonalpha} also becomes relevant. For $\Gamma/\mathcal{H}\vert_{\TR} \gg 1$ the asymmetry decreases exponentially, resulting in a maximum around $\Gamma/\mathcal{H}\vert_{\TR} \sim \mathcal{O}(1)$, depending on $\alpha_3$.

Let us provide one numerical example to illustrate the size of the underlying Yukawa couplings. Setting
\begin{align}
F_1 &= \ii F_2 = 0.25\,, \, G_1 = G_2 = K_1 = -K_2 = 2.5\,,\\
\TR &= \unit[10^{14}]{GeV}\,, \, M_1 = \unit[10^{15}]{GeV}\,, \, M_2 = \unit[10^{17}]{GeV}\nonumber
\end{align}
gives the Boltzmann-equation parameters
\begin{align}
\alpha_1 = \alpha_2/100 = \alpha_3 \simeq 2 \times 10^{-5}\,, \quad \epsilon \simeq 2 \times 10^{-10}\,,
\end{align}
with interaction strength $\Gamma/\mathcal{H}\simeq 4$. This is near the efficiency maximum in Fig.~\ref{fig3a} and thus gives the correct baryon asymmetry.
The naive one-loop contribution to neutrino masses $\mathcal{O}( F^* m_d G/16 \pi^2)$ comes out to $\unit[2\times 10^{4}]{eV}$ here, which -- one might argue -- requires a fine-tuned cancellation to eventually end up with sub-eV neutrino masses. While this finetuning can be softened in other areas of parameter space, our restriction to $\TR > \unit[3\times 10^{13}]{GeV}$ makes it difficult to push the Yukawa couplings to small-enough values to alleviate finetuning altogether. We will study smaller reheating temperatures and more suppressed Yukawa couplings in future work.

For large $\nu_R$ interaction rates, $\Gamma/\mathcal{H}\vert_{\TR} \gg 1$, the $\nu_R$ density $\Sigma_{\nu_R}$ starts to saturate the equilibrium value. Since the single-flavor $\nu_R$ is decoupled from the SM bath, this essentially massless fermion contributes to the radiation density in the early universe, usually parametrized via the effective number of neutrinos, $N_\text{eff}$ (see \eg~Ref.~\cite{Heeck:2023lxl}):
\begin{align}
    \Delta N_\text{eff}&\simeq 0.047\, \bigg(\frac{106.75}{g_*}\bigg)^{4/3} \frac{\Sigma_{\nu_R}(\infty)}{\Sigma_{\nu_R}^\eq} \label{eq:Neff}\\
    &\simeq 0.047\, \bigg(\frac{106.75}{g_*}\bigg)^{4/3} \bigg(1-\exp\bigg[-\frac{\Gamma}{3\mathcal{H}}\bigg\vert_{\TR}\bigg]\bigg) \,.\nonumber
\end{align}
We show $\Delta N_\text{eff}$ together with the $\nu_R$ asymmetry in Fig.~\ref{fig3}. 
Even one fully thermalized $\nu_R$ only contributes $\Delta N_\text{eff} \simeq 0.047$ given the high decoupling temperature, which is well within the currently allowed range~\cite{Planck:2018vyg} and even below the upcoming CMB-S4 sensitivity of  $\Delta N_\text{eff}\sim 0.06$~\cite{Abazajian:2019oqj, Abazajian:2019eic}.

Overall, Dirac leptogenesis via scattering is an efficient mechanism to explain the matter--antimatter asymmetry of our universe in models with Dirac neutrinos, without ever violating $B-L$.
In the single-flavor -- effectively one-generational -- case studied above, this extends the viable parameter space of all Dirac-leptogenesis models of Ref.~\cite{ Heeck:2023lxl} that contain at least three different Yukawa matrices, \ie~models \emph{b} and \emph{c}, into the region $\TR \ll M_X$.
Within our approximations, the models with only two Yukawa matrices would yield a vanishing asymmetry~\cite{Blazek:2023dow}. However, the introduction of flavor effects could change this conclusion and allow for all Dirac leptogenesis models to work in the low-reheating regime. The flavor becomes important when we allow $X$ to couple to more than one generation and when we lower the reheating temperature below $\unit[3\times 10^{13}]{GeV}$, where sphalerons and SM Yukawa interactions become relevant. $X$ couplings to two or even all three $\nu_R$ will also increase the generated $\Delta N_\text{eff}$ from Eq.~\eqref{eq:Neff} by a factor two or three, respectively, both of which are currently allowed but within reach of  CMB-S4~\cite{Abazajian:2019oqj, Abazajian:2019eic} unless $\Gamma/\mathcal{H}\vert_{\TR}\ll 1$. At least part of the Dirac leptogenesis parameter space is hence testable through $\Delta N_\text{eff}$. A detailed discussion of flavor effects is left for future work.

\section{Conclusion}
\label{sec:conclusions}

Neutrino oscillations have proven neutrinos to be massive particles, but without any indication of whether they are Dirac or Majorana particles. While we anxiously await an experimental resolution of this question through studies of neutrinoless double-beta decay, it behooves us to investigate possible impacts of \emph{Dirac neutrinos}. Compared to Majorana neutrinos, the connection between neutrino mass and matter--antimatter asymmetry has hardly been explored, even though Dirac leptogenesis~\cite{Dick:1999je} beautifully employs the neutrino-mass smallness to hide a lepton asymmetry in the decoupled $\nu_R$ bath, letting sphalerons create a baryon asymmetry in a $B-L$-conserving universe. Similar to standard leptogenesis, this is easily accomplished through the out-of-equilibrium decays of new heavy particles~\cite{Dick:1999je, Heeck:2023lxl}.

In this paper, we have demonstrated that the non-thermalization of the $\nu_R$ is sufficient to satisfy Sakharov's conditions \textit{without} the need for heavy mediator decays. This makes it possible to obtain the observed baryon asymmetry purely from scatterings without ever producing the mediators. This is relevant for inflationary scenarios with reheating temperatures below the mediator mass scale.

In the single-flavor limit employed here, a non-vanishing asymmetry requires the existence of at least three different scattering states in the considered $2\to2$ processes, which we exemplified by considering heavy leptoquarks mediating the interactions $\nu_R d_R\leftrightarrow LQ$, $\nu_R d_R \leftrightarrow e_R u_R$, and $e_R u_R \leftrightarrow LQ$. While the source term for the $\Delta_{\nu_R}$ vanishes in the Boltzmann equations, the existence of wash-in processes does enable the $\nu_R$ asymmetry generation, provided there exists at least one other conserved charge, taken to be right-handed electron number in our main-text example.

We show that sufficiently large asymmetries can be achieved in a wide range of parameter space with a maximal value reached for intermediate interaction strength that lie between the freeze-in regime (very weak couplings) and the semi-thermalized regime (larger couplings). 
The viable window is limited toward smaller couplings as the $\nu_R$ abundance becomes too small to provide large $\Delta_{\nu_R} $ and towards larger couplings by the diminishing deviation from thermal equilibrium of $\nu_R$. The multi-generational scenario can be probed by an enhanced $N_\text{eff}$: while evading current constraints, CMB-S4 experiments are expected to test the semi-thermalized regime as well as the maximum asymmetry scenario. 

\section*{Acknowledgements}

This work was partly supported by the National Science Foundation under Grant No.~PHY-2210428. J.~Heisig acknowledges support from the Alexander von Humboldt Foundation via the Feodor Lynen Research Fellowship for Experienced Researchers and the Feodor Lynen Return Fellowship. T.~Bla\v{z}ek, P.~Mat\'ak, and V.~Zaujec were supported by the Slovak Grant Agency VEGA, project No. 1/0719/23.

\bibliographystyle{bibstyle_mod}
\bibliography{bib.bib}

\end{document}